\documentclass[preprint,proceedings]{rmaa}

\newcommand{\beq}{\begin{equation}}
\newcommand{\eeq}{\end{equation}}
\newcommand{\beqn}{\begin{eqnarray}}
\newcommand{\eeqn}{\end{eqnarray}}
\newcommand{\non}{\nonumber}
\newcommand{\pa}{\partial}

\SetYear{2004}
\SetConfTitle{The Environment and Evolution of Binary and Multiple Stars}

\title{The Modified Restricted Three Body Problems}

\author{Ing-Guey Jiang\altaffilmark{1} and Li-Chin Yeh\altaffilmark{2}}

\altaffiltext{1}{National Central Univ., Taiwan. jiang@astro.ncu.edu.tw}
\altaffiltext{2}{National Hsinchu Teachers College, Taiwan.\\ lcyeh@bsd.nhctc.edu.tw }

\shortauthor{Jiang and Yeh}
\shorttitle{Three Body Problems}

\fulladdresses{
\item Ing-Guey Jiang: Institute of Astronomy, National Central University,
 Taiwan (\email{jiang@astro.ncu.edu.tw}).
\item Li-Chin Yeh: Department of Mathematics, National Hsinchu Teachers 
 College, Hsin-Chu, Taiwan\\ (\email{lcyeh@bsd.nhctc.edu.tw}).}

\listofauthors{Ing-Guey Jiang, Li-Chin Yeh }
\indexauthor{Jiang, I.-G.}
\indexauthor{Yeh, L.-C.}

\addkeyword{Three-body problem}

\abstract{The restricted three body problem is well-known and very important 
for dynamics of binary, multiple stars and also planetary systems. We extend 
the classical version of this problem to the situation that there are some 
external forces from the belt. We find that both the equilibrium points and 
solution curves become quite different from the classical case. We also 
determine the values of Lyapunov Exponent for some important orbits.}

\resumen{

El problema restringido de los tres cuerpos es importante en la din\'amica
de las estrellas dobles y multiples y de los sistemas planetarios.  Extendemos
la versi\'on clásica de este problema a una situaci\'on que incluye
un anillo.  Encontramos puntos de equilibrio y curvas muy distintas a las
del caso cl\'asico.  Calculamos el valor del exponente de Liapunov para algunas
\'orbitas. } 

\addkeyword{Three body problem}

\begin{document}
\maketitle 

\section{Introduction}

The three body problem is one of the most important problems of 
celestial mechanics and has been analytically and numerically studied 
for centuries. In addition to that, 
three body interaction also plays an essential role for dynamics 
of binary and multiple stars. Please see Valtonen (2004) and 
Dvorak (2004) and also their references. 

On the other hand, because 
there are Asteroid Belt and Kuiper Belt for the Solar system,
discs of dust for extrasolar planetary systems and also circumbinary
rings for binary systems, these belt-like structure should 
influence the dynamical evolution of these systems. For instance, 
Jiang \& Ip (2001) show that the origin of orbital elements of  
the planetary system of upsilon Andromedae
might be influenced by the belt interaction initially.
Moreover, Yeh \& Jiang (2001) studied the orbital migration of scattered 
planets. They completely classify the parameter space and solutions
and conclude that the eccentricity always increases if the planet, which 
moves on circular orbit initially, is scattered to migrate outward.
Thus, the orbital circularization must be important for scattered planets
if they are now moving on nearly circular orbits.

Therefore, Jiang \& Yeh (2003) did some analysis on the solutions for 
dynamical systems of planet-belt interaction. In this paper, we further study 
the effect of belts for dynamical evolution of a binary system.

\section{The Model}

We consider the motion of a test particle influenced by 
the gravitational force from the central binary and the circumbinary belt.
The circumbinary belt also provides the frictional force for the test particle.

We assume that two masses of the central binary 
are $m_1$ and $m_2$ and choose the unit of mass to make
$G(m_1+m_2)=1$. If we define that $$\bar{\mu}=\frac{m_2}{m_1+m_2},$$
then the two masses are
$\mu_1=Gm_1=1-\bar{\mu}$ and $\mu_2=Gm_2=\bar{\mu}$.
The separation of central binary is set to be unity
and $\mu_1=\mu_2=0.5$ for all numerical results in this paper.

The equation of motion of this problem is 
(Murray \& Dermott 1999)
\beq\left\{
\begin{array}{ll}
&\frac{dx}{dt}=u·\\
&\frac{dy}{dt}=v  \\
&\frac{du}{dt}=2v-\frac{\pa U^{\ast}}{\pa x}-\frac{\pa V}{\pa x}+
f_{\alpha x}  \\
& \frac{dv}{dt}=-2u-\frac{\pa U^{\ast}}{\pa y}-\frac{\pa V}{\pa y}
+f_{\alpha y},
\end{array} \right. \label{eq:3body1} 
\eeq
where the potential $U^{\ast}$ is 
\beq
U^{\ast}=-\frac{1}{2}(x^2+y^2)-\frac{\mu_1}{r_1}-\frac{\mu_2}{r_2},
\label{eq:u_ast}
\eeq
 $r_1=\sqrt{(x+\mu_2)^2+y^2}$ and $r_2=\sqrt{(x-\mu_1)^2+y^2}$.
$V$ is the potential from the belt.
The belt is a annulus with inner radius $r_{i}$ and
outer radius $r_{o}$, where  $r_{i}$ and $r_{o}$ are assumed to be 
constants. We arbitrary set $r_{i}=0.2$ and $r_{o}=1.0$ for all results
in this paper.

The density profile of the  belt is $\rho(r)=c/r^p$, where $r=\sqrt{x^2+y^2}$,
$c$ is a constant 
completely determined by the total mass of the belt and $p$ is a natural 
number. In this paper, we set $p=2$ for all numerical results. Hence, 
for $p=2$, the total mass of the belt is 
\beq
M_{b}=\int^{2\pi}_{0}\int^{r_o}_{r_i}\rho(r')r'dr'd\phi =2\pi c
(\ln r_o-\ln r_i). 
\eeq

The gravitational force $f_b$ from the belt is 

\beq
f_b(r)=-\frac{\pa V}{\pa r}= 
-2\int^{r_o}_{r_i}\frac{\rho(r')r'}{r}\left[\frac{E}{r-r'}+
\frac{F}{r+r'}\right] dr',\label{eq:fb}
\eeq
where $F(\xi)$ and $E(\xi)$ are elliptic 
integral of the first kind and the second kind.
Hence, 
\beq\left\{
\begin{array} {ll}
&-\frac{\pa V}{\pa x}= f_b\frac{x}{r}\\ &-\frac{\pa V}{\pa x}= f_b\frac{y}{r}, 
\end{array} \right.\label{eq:pav}
\eeq
where $f_b$ is in Eq. (\ref{eq:fb}).

The frictional force should be proportional to the surface density of 
the belt  and the velocity of the particle. 
In the x direction, the frictional force is 
\beq
f_{\alpha x} =-\alpha \rho(r) \frac{dx}{dt} \label{eq:fbx} 
\eeq
and in the y direction, the frictional force is 
\beq
f_{\alpha y} = -\alpha \rho(r) \frac{dy}{dt}, \label{eq:fby}
\eeq
where $\alpha$ is the frictional parameter.

We substitute Eq. (\ref{eq:u_ast}) and Eq. (\ref{eq:fb})-(\ref{eq:fby}) into 
Eq.(\ref{eq:3body1}) and have the following system:

\beq \left\{
\begin{array}{ll}
&  \frac{dx}{dt}=u \\
&  \frac{dy}{dt}=v  \\
& \frac{du}{dt}=2v +x-\frac{\mu_1(x+\mu_2)}{r_1^3}-\frac{\mu_2(x-\mu_1)}
{r_2^3}-\frac{2x}{r^2}   \\
& \times  \int^{r_o}_{r_i}\rho(r')r'
\left[\frac{E}{r-r'}+\frac{F}{r+r'}\right]dr'
-\alpha\rho(r) u  \\
& \frac{d v}{dt}=-2u+y-\frac{y\mu_1}{r_1^3}-\frac{y\mu_2}{r_2^3}
-\frac{2y}{r^2}   \\
& \times  \int^{r_o}_{r_i}\rho(r')r'\left[\frac{E}{r-r'}
+\frac{F}{r+r'}\right]dr'-\alpha\rho(r)v.   
\end{array}  \right. \label{eq:3body2}
\eeq
\section{Equilibrium Points}

The equilibrium points $(x_e,y_e)$ of System (\ref{eq:3body2}) satisfy the 
following equations 
\beqn
& & f(x,y)\equiv x-\frac{\mu_1(x+\mu_2)}{r_1^3}-\frac{\mu_2(x-\mu_1)}
{r_2^3} \non \\
 & &  -\frac{2x}{r^2}\int^{r_o}_{r_i}\rho(r')r'
\left[\frac{E}{r-r'}+\frac{F}{r+r'}\right]dr'=0,  \\
& &g(x,y)\equiv y-\frac{y\mu_1}{r_1^3}-\frac{y\mu_2}{r_2^3} -\frac{2y}{r^2}
\non \\
& & \int^{r_o}_{r_i}\rho(r')r'\left[\frac{E}{r-r'}
+\frac{F}{r+r'}\right]dr'=0.
\eeqn
In Fig.1, we plot the curves of $f(x,y)=0$ (full circles) 
and $g(x,y)=0$ (open triangles) for different
values of  $M_b$. Equilibrium points $(x_e,y_e)$ are intersection of these 
two. In Fig.1 (a), we set $M_b=0$, so there is no influence 
from the belt. We find there are five  usual Lagrangian points, $L_1$, $L_2$,
$L_3$, $L_4$ and $L_5$ for this case. 
In Fig. 1 (b), $M_b=0.15$ and we still have five usual 
Lagrangian points. In addition to that, there are two new equilibrium points 
 in the upper half-plane and another two  in the lower half-plane.
In Fig. 1 (c)-(d), we set $M_b=0.3$ and $M_b=0.5$ individually.
We also find that in the upper half-plane, there are two new 
equilibrium points, $F_a$ and $F_b$.

\begin{figure}[!t]
  \includegraphics[width=\columnwidth]{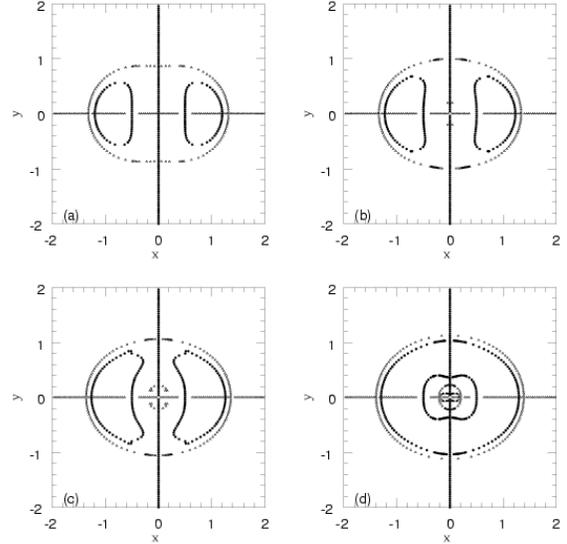}
  \caption{The curves of $f(x,y)=0$ and  $g(x,y)=0$ (see the text for details).}
  \label{Figure_fix}
\end{figure}

\section{Lyapunov Exponent}

Since we have discovered two new equilibrium points near $L_4$ (and another 
two near $L_5$), it would be interesting to investigate the orbital behavior
around these new equilibrium points $F_a$ and $F_b$. For a complicated system
like ours, it is difficult to rigorously prove if the orbits are chaotic 
near $F_a$ and $F_b$. Nevertheless, we use the calculation of Lyapunov Exponent
for some orbits whose initial conditions are chosen to be close to 
$F_a$ and $F_b$ to understand how sensitively  
dependent on the initial conditions for these orbits. 
This is in fact one of the most important methods to study 
chaotic systems. 
We follow Wolf et al. (1985) to calculate the values of Lyapunov Exponent
numerically.
To check if our calculation is correct, we have reproduced the results
of a given system in their paper.

In general, 
the larger value of Lyapunov Exponent  means 
more sensitively  
dependent on the initial conditions. 
We choose the initial conditions of orbits to be close to the  
equilibrium points $F_a$, $F_b$, $L_4$ and $L_2$ individually. Thus,
there are 4 different initial conditions for the orbital calculations.
To understand the effect of the belt with different mass, we did
calculations for 4 different masses of the belt ($M_b=0,0.15,0.3,0.5$) 
for each chosen initial conditions. 
Although there are no equilibrium points $F_a$, $F_b$ when there is no
belt ($M_b=0$), and the locations of equilibrium point $F_a$, $F_b$,
$L_4$ and $L_2$ would be slightly different for different masses of the belt,
we still call the initial condition 
$(x,y,u,v)=(0.01,0.0225,0,0)$ Initial Condition $F_a$, 
$(x,y,u,v)=(0.01,0.06,0,0)$  Initial Condition  $F_b$, 
$(x,y,u,v)=(0.01,1,0,0)$ Initial Condition  $L_4$ and
$(x,y,u,v)=(1.35,0,0,0)$ Initial Condition  $L_2$.

Fig.2(a)-(d) are the results of Lyapunov Exponent for  
Initial Condition $F_a$, $F_b$, $L_4$ and $L_2$ individually. 
There are 4 curves in each panel of
Fig.2, where solid curve is the result of $M_b=0$, 
dotted curve is the result of $M_b=0.15$,
dashed curve is the result of $M_b=0.3$ and
long dashed curve is the result of $M_b=0.5$.
It is obvious that the values of Lyapunov Exponent for  Initial Condition 
$F_a$, $F_b$ is much larger than the ones for Initial Condition  $L_4$ and
$L_2$. 
From Panel (a) and (b),
we can also see that the Lyapunov Exponent for $M_b=0.5$ and 
$M_b=0.3$ are larger than the values for $M_b=0.15$ and 
$M_b=0$. 
Interestingly, for orbits with 
Initial Condition  $L_4$,
the values of Lyapunov Exponent 
 are slightly larger for $M_b=0$ as we can see in Fig.2(c). 
In general, their
values are small for both Initial Condition $L_4$ and $L_2$. 
The values of  Lyapunov Exponent
for orbits with Initial Condition $L_2$ 
approach to 0 when $t$ tends to infinite. The orbits 
are obviously not chaotic for this case.

\begin{figure}[!t] 
  \includegraphics[width=.95\columnwidth]{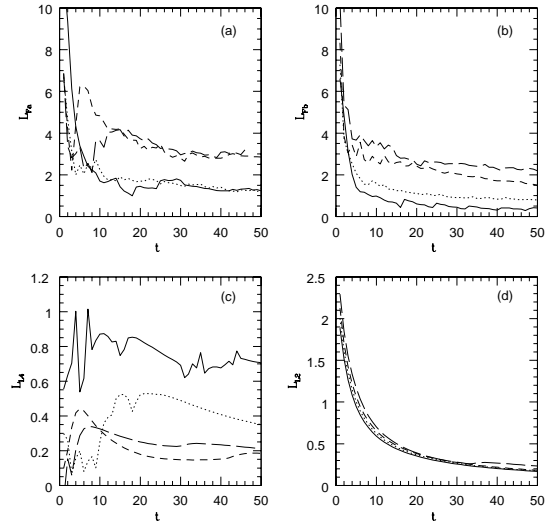}
  \caption{Lyapunov Exponent (see the text for details).}
  \label{Figure1}
\end{figure}

\section{Orbits}

In this section, we will discuss all the orbits whose results of 
Lyapunov Exponent have been shown and discussed in last section. 

Fig.3, 4, 5 and 6 are the orbits on $x-y$ plane for  
Initial Condition $F_a$, $F_b$, $L_4$ and $L_2$. There are 4 panels for 
each figure. Panel (a) is the result when there is no belt, i.e. $M_b=0$,
panel (b) is the result when  $M_b=0.15$,
panel (c) is for  $M_b=0.3$  and
panel (d) is the result for  $M_b=0.5$.

If one looks at all these 4 figures of orbits at the same time, 
one can immediately understand that it seems the orbits with Initial 
Condition $F_a$ and $F_b$ are much more chaotic than the orbits
with Initial Condition $L_4$ and $L_2$.
This impression is completely consistent with the one we get from the 
values of Lyapunov Exponent.

To compare  Fig.3(a) with Fig.4(a), we found that 
the orbits are similar for Initial 
Condition $F_a$ and $F_b$. However, from the comparison between 
Fig.3(b) and Fig.4(b), we found that 
the orbits are quite different for Initial Condition $F_a$ and $F_b$.
These two comparisons show that the existence of belt
does make the orbits become more sensitive to the initial conditions.

\begin{figure}[!t]\centering
  \includegraphics[width=.9\columnwidth]{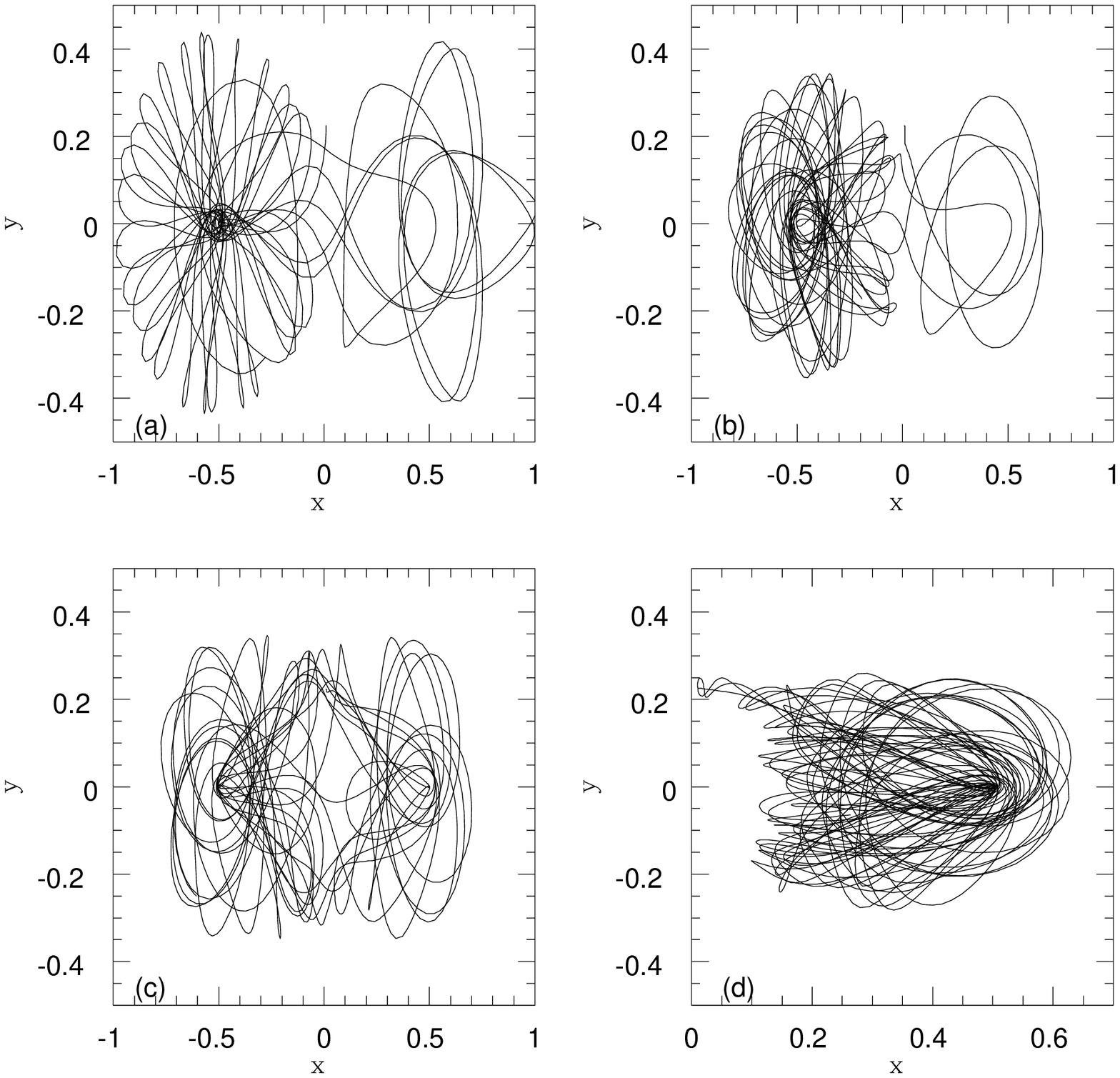}
  \caption{The orbits with Initial Condition $F_a$.} 
  \label{Figure2}
  \includegraphics[width=.9\columnwidth]{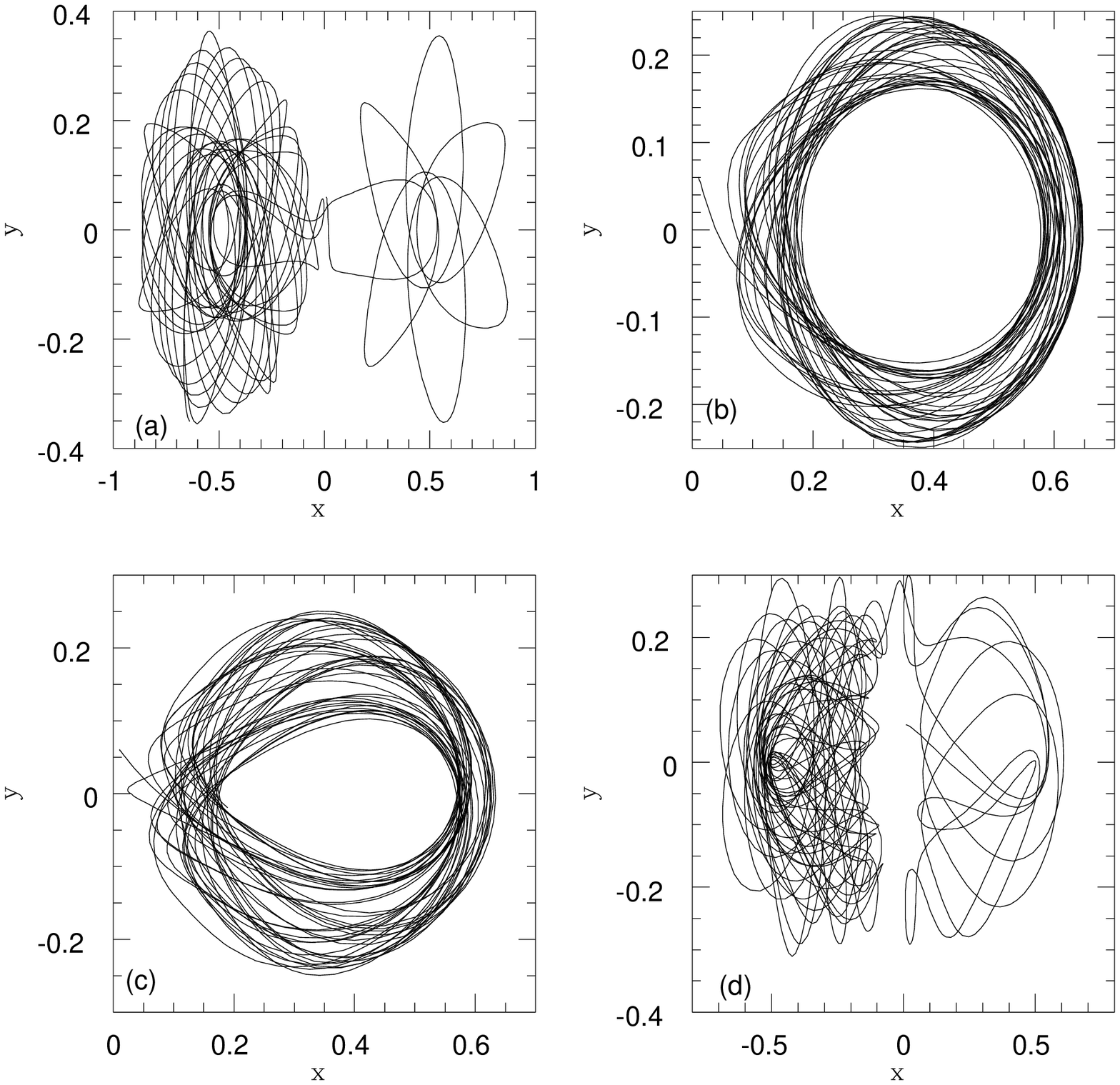}
  \caption{The orbits with Initial Condition $F_b$.} 
  \label{Figure3}
\end{figure}

\begin{figure}[!t]\centering
  \includegraphics[width=.9\columnwidth]{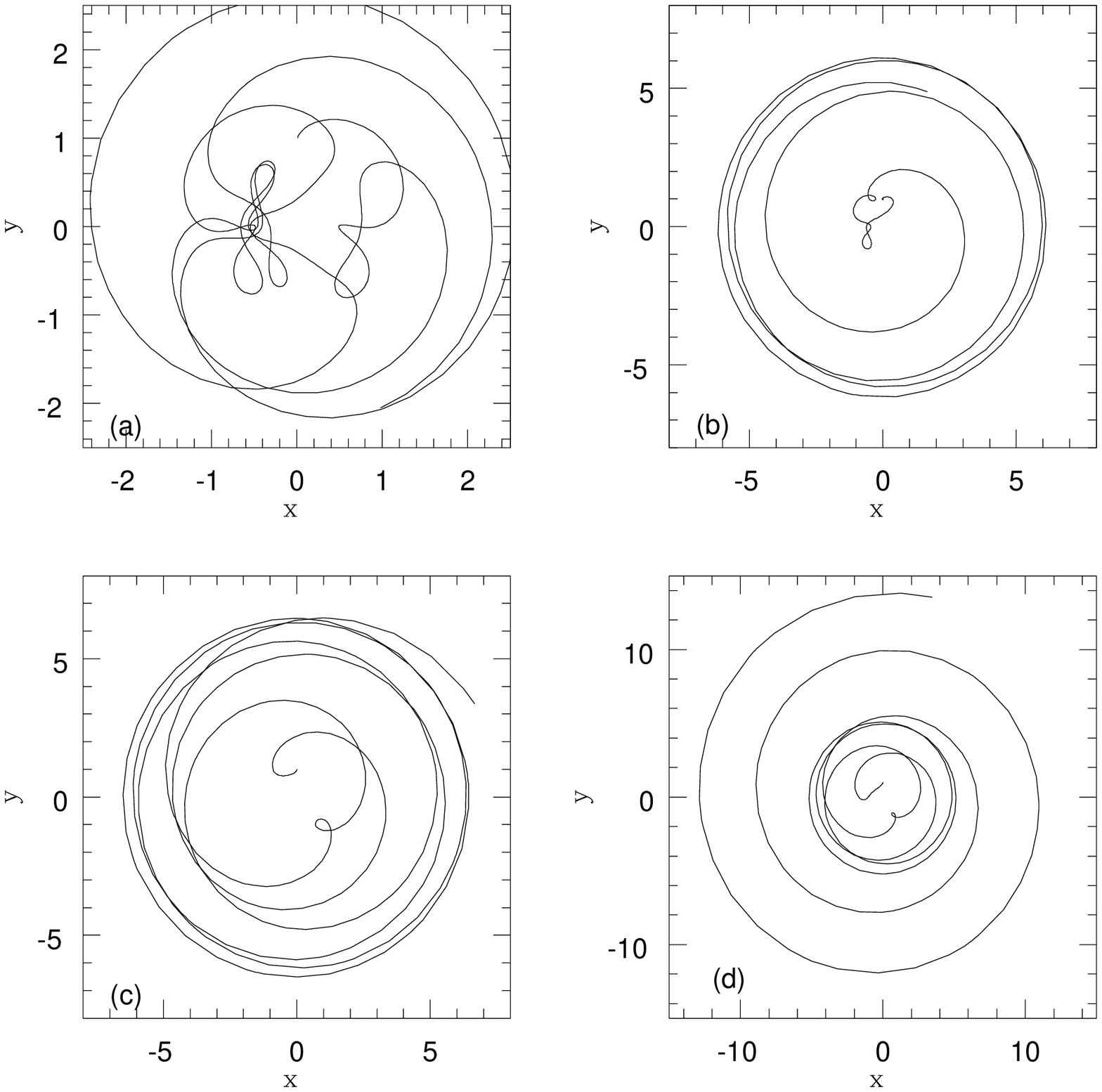}
  \caption{The orbits with Initial Condition $L_4$.}
  \label{Figure4}
  \includegraphics[width=.9\columnwidth]{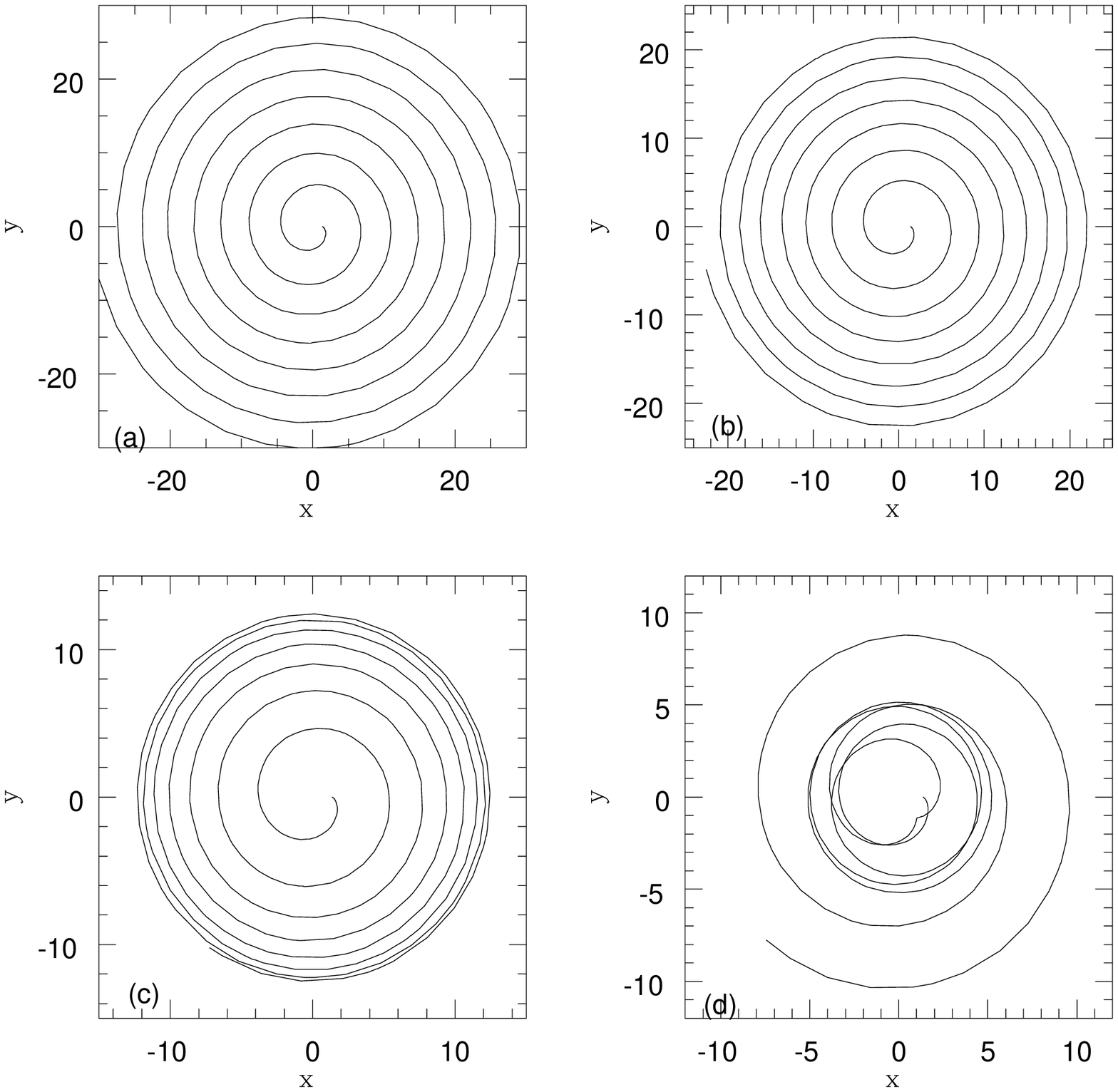}
  \caption{The orbits with Initial Condition $L_2$.}
  \label{Figure5}
\end{figure}

\section{Concluding Remarks}

We have provided the equations for a model which modifies the
classical restricted three body problem by including the influence from 
a belt around the central binary. We found that, in addition to the usual
Lagrangian points, there are two new equilibrium points, which we
call $F_a$ and $F_b$ around $L_4$ (similarly, there are another two new
equilibrium points close to $L_5$).

To study the orbits around these new equilibrium points, we 
calculate the values of 
Lyapunov Exponents for orbits with 4 different initial conditions, 
which are close to $F_a$ and $F_b$, $L_4$ and $L_2$ individually.
We found that the belt makes the system even more
sensitive to the initial conditions for the
orbits with Initial Condition $F_a$ and $F_b$ but does not 
make too much different for the orbits 
with Initial Condition $L_4$ and $L_2$.
Because the equilibrium points $F_a$ and $F_b$ happen to be near inner
part of the belt and Lagrangian points $L_4$ and $L_2$ happen to be 
around or out of outer part of the belt in our system, it seems that 
the orbits near inner part of the belt might be more unpredictable
than the ones around 
outer part. 

\vspace*{.4\baselineskip}
\adjustfinalcols

\end{document}